\documentclass{elsart1p}
\usepackage{graphicx}
\def\lsim{\mathrel{\rlap{\lower4pt\hbox{\hskip1pt$\sim$}}
    \raise1pt\hbox{$<$}}}         
\def\gsim{\mathrel{\rlap{\lower4pt\hbox{\hskip1pt$\sim$}}
    \raise1pt\hbox{$>$}}}         
\usepackage{amssymb}
\begin{document}
\begin{frontmatter}
%
%
%
\title{CN-Cycle Neutrinos and Solar Metalicity}
%
%
\author{W. C. Haxton$^\dagger$ and A. M. Serenelli$^{\dagger \dagger} $}
\address{$^\dagger$Inst. for Nuclear Theory and Dept. of Physics, 
  University of Washington, Seattle, WA 98195 USA \\
  $^{\dagger \dagger}$Max-Planck Institute f\"ur Astrophysik, Karl-Schwarzschild-Str. 1, 85748 Garching, Germany}
\begin{abstract}
 We discuss the current Standard Solar Model conflict between helioseismology
 and photospheric abundances, a speculation that connects this anomaly to
 formation of the gaseous giant planets, and a possible neutrino measurement to
 directly test solar core metalicity.
\end{abstract}
\end{frontmatter}
%
\section{Helioseismology and Photospheric Abundances}

This talk summarizes some recent work \cite{HS} on a 
discrepancy in the Standard Solar Model (SSM) -- a conflict between helioseismology 
and the new metal abundances that emerged from improved modeling of
the photosphere --  and on a possible connection to a key SSM assumption, that the early
Sun was chemically homogeneous due to its passage through the fully convective Hayashi phase.  
We suggest a speculative mechanism -- planetary formation -- that could invalidate this
assumption, and an opportunity, CN-cycle neutrinos, for
independently determining the Sun's central metalicity. 

The SSM assumes local hydrostatic equilibrium and proton burning by the pp chain and CN cycle,
with the latter accounting for about 1\% of energy generation.  The Sun's evolution from 
zero-age on the main sequence is constrained by various boundary conditions (initial mass, present luminosity, etc.),
including the initial composition.  Assuming a homogeneous proto-Sun, the initial core
metalicity (Z) is fixed to today's surface abundances under the assumption that these have 
changed little over the past 4.6 b.y. of solar evolution, while the He/H ratio is adjusted to
produce the correct modern luminosity.  Small corrections due to diffusion of heavy elements
are made in the model.

Photospheric absorption lines are the only practical way to fix the abundances of certain 
volatile elements such as C, N, and O.  Metals influence the SSM through bound$\leftrightarrow$free transitions
that affect the opacity, with O and Ne being important for temperatures characteristic of the upper radiative zone.  Until recently, metalicities
determined from interpretations of photospheric absorption lines, e.g.,  the 1998
work of Grevesse and Sauval \cite{GS98}, led to SSM sound speed profiles
that agreed with helioseismology.

These earlier line analyses were based on 1D models of the photosphere, despite known stratification, convection, and inhomogeneities.   To address these deficiencies, parameter-free 3D
models were developed.  These more complete models markedly improved line shapes and  the consistency of line sources.  The new analyses \cite{AGS05}, however, led to a 
significant reduction in Z, 0.0169 $\rightarrow$ 0.0122, altering SSM sound speeds and 
destroying the once good agreement between helioseismology and the SSM (see Fig. 1).

\begin{figure}
\centering
\includegraphics[width=10cm]{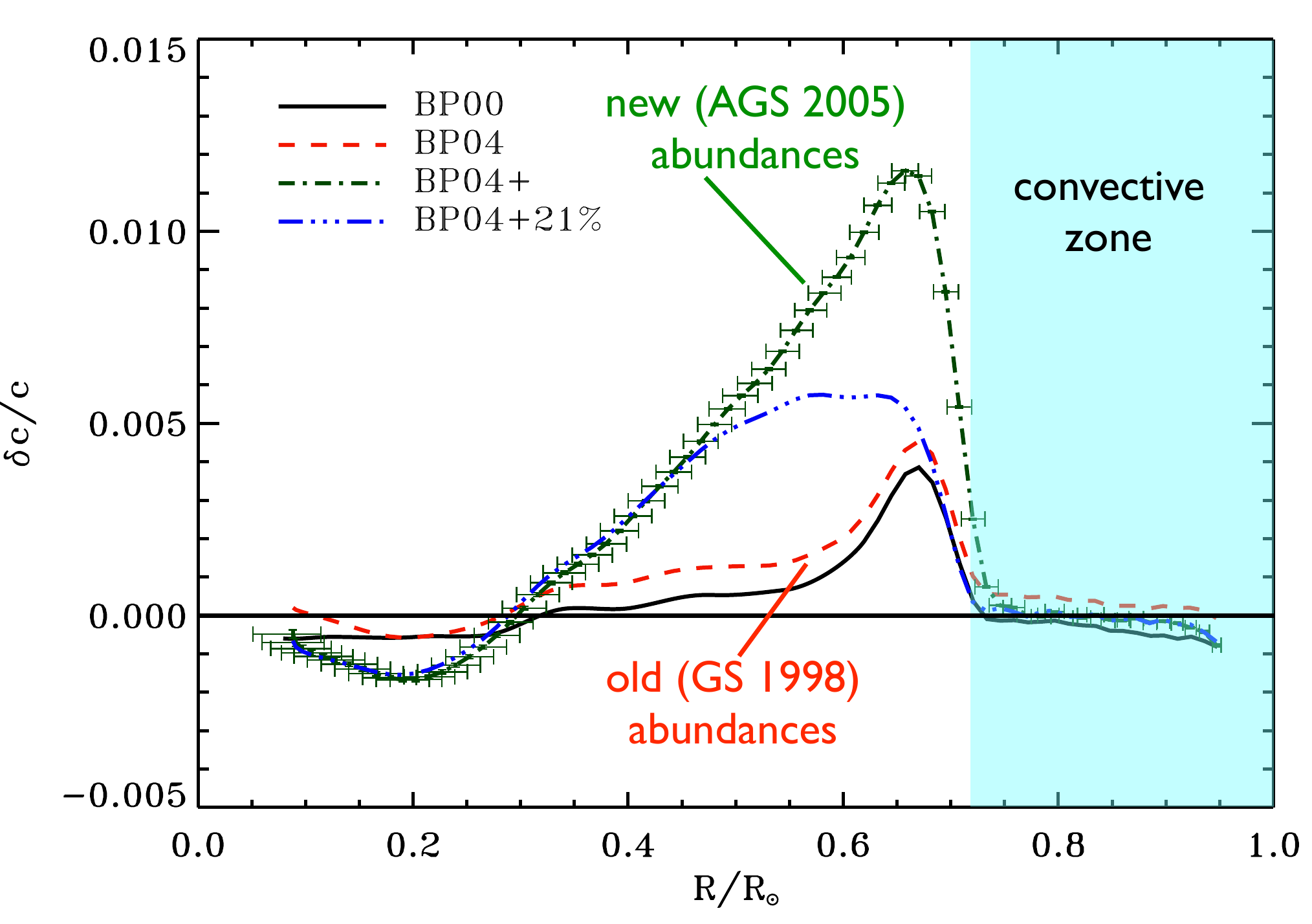}
\caption{The sound speed discrepancies for GS98 and AGS05 abundances.  See \cite{HS} for discussion.}
\end{figure}

The reduced Z also affects the SSM $^8$B neutrino flux, due to the
sensitivity of this prediction to core temperature.  The change from GS98 \cite{GS98} to
AGS05 \cite{AGS05} abundances lowers the $^8$B flux prediction from 5.95 to
4.72 $\times$ 10$^6$/cm$^2$s.  The 391-day SNO NCD-phase result is
5.54 $\pm$ 0.32 $\pm$ 0.35 $\times$ 10$^6$/cm$^2$s \cite{SNO}.

\section{Metals in the Early Solar System}
The convective zone extends over the outer 30\% of the Sun by radius and contains about
3\% of the Sun's mass.  The change from GS98 to AGS05 abundances
lowers the total metal content of this zone by  50 $M_\oplus$.  Interestingly, the one known
example of large-scale metal segregation in the solar system, the formation of the gaseous
giant planets, concentrates a similar amount of metal, $\sim$ 40-90 $M_\oplus$, depending
on modeling uncertainties.  The conventional picture (see \cite{HS} for references) places
planetary formation late in the development of the solar nebula, when the last few percent 
of the gas has formed into a disk,  with metal-rich grains and ice concentrated in the disk's
midplane.  In the core accretion model, midplane interactions allow rocky planetary cores to grow until  $\sim$ 10$M_\oplus$, after which the gravitational potential is sufficient to capture gas.  Envelope formation is thought to be rapid, requiring perhaps as little at 1My.  This process produced
metal enrichments of Jupiter and Saturn of $\sim$ 3-7 \cite{guillot}.

The process of planetary formation, by scrubbing metals from an initially homogeneous gas cloud, 
would produce enough metal-depleted gas to dilute the convective zone.  This could lead to a two-zone Sun -- a core higher in  Z than the surface -- contradicting 
a key SSM assumption and possibly accounting for the apparent discrepancy between helioseismology
and photospheric abundances.

This conjecture passes some simple tests connected with the total budget of metals and the
total mass of gas that a Jupiter could perturb gravitationally during planetary formation.  It
requires 1) planetary formation to occur after the Sun developed a radiative core (separating the interior from the exterior) and
2) deposition of a significant fraction of the metal-poor gas onto the Sun. These assumptions do not appear  unreasonable \cite{HS}.

There are several variables in this picture, including the amount of gas processed, the efficiency of the
fractionation, whether the fractionation affects all elements equally, and the dynamics of the depleted
gas.  The constraints include the photospheric abundances and partial abundances  for
Jupiter and Saturn, determined from Galileo, Cassini, and subsequent modeling \cite{guillot}.
One would need to explore this parameter space to test whether this scenario could
quantitatively account for observations.

\section{Can Neutrinos Help?}
It would be helpful to test the SSM assumption of a homogenous
Sun by directly comparing abundances on the surface and in the core.  We noted earlier
that the $^8$B neutrino flux responds to changes in metalicity due to the influence of metals on core temperature.  But the change is modest and not characteristic: many of the 19 parameters
of the SSM can be adjusted to produce similar core temperature changes.  Changes in
fluxes due to parameter variations that alter core temperature will be termed ``environmental."

But CN solar neutrino sources have a linear dependence on core metalicity, in addition to
the environmental sensitivity.  The BPS08(GS) SSM \cite{BPS} predicts a modest 0.8\% CN-cycle
contribution to solar energy generation but measurable neutrino fluxes, e.g.,
\begin{equation}
^{15}O(\beta^+)^{15}N~~~E_\nu \lsim 1.732 \mathrm{~MeV}~~~\phi=(2.20^{+0.73}_{-0.63}) \times 10^8 \mathrm{cm}^2 \mathrm{s}.
\end{equation}
Because the CN and $^8$B neutrinos have a similar dependence on the core temperature,
environmental uncertainties (solar age, opacity, luminosity,...) produce correlated changes
in these fluxes.  This correlation (see Fig. 2) allows one to use the 
measured $^8$B flux  \cite{SNO,SK} to 
largely eliminate environmental uncertainties affecting the CN flux, yielding 
\begin{eqnarray}
{\mathrm{R^{SNO+}(CN)} \over \mathrm{R^{SSM}(CN)}} &=& {\mathrm{X(C+N)} \over \mathrm{X^{SSM}(C+N)}}\left( \mathrm{{R^{SK}(^8B)} \over \mathrm{R^{SSM}(^8B)}}\right)^{0.828} \nonumber \\
&\times&  \left[1 \pm 0.03(\mathrm{SK}) \pm 0.026 (\mathrm{res~env})\pm0.049 (\mathrm{LMA}) \pm 0.071 (\mathrm{nucl}) \right] 
\end{eqnarray}
The ratio of the CN-neutrino rate R measured in a future deep scintillator detector (e.g.,
SNO+ \cite{SNO+}, Borexino \cite{Borexino}, or Hanohano \cite{Hanohano}) to that calculated in the SSM appears on the left side.
The quantity of interest,
the ratio of the  primordial core C+N metalicity X to the SSM value, appears as the first term on
the right.  The proportionality between these ratios
can be expressed in terms of the ratio of the measured and SSM rates for Super-Kamiokande (SK).  Residual
uncertainties include the  SK $^8$B measurement error,  remaining environmental dependences (after use of the SK constraint), neutrino
oscillation parameters, and nuclear cross sections.  Further details are given
in \cite{HS}.

\begin{figure}
\centering
\includegraphics[width=10cm]{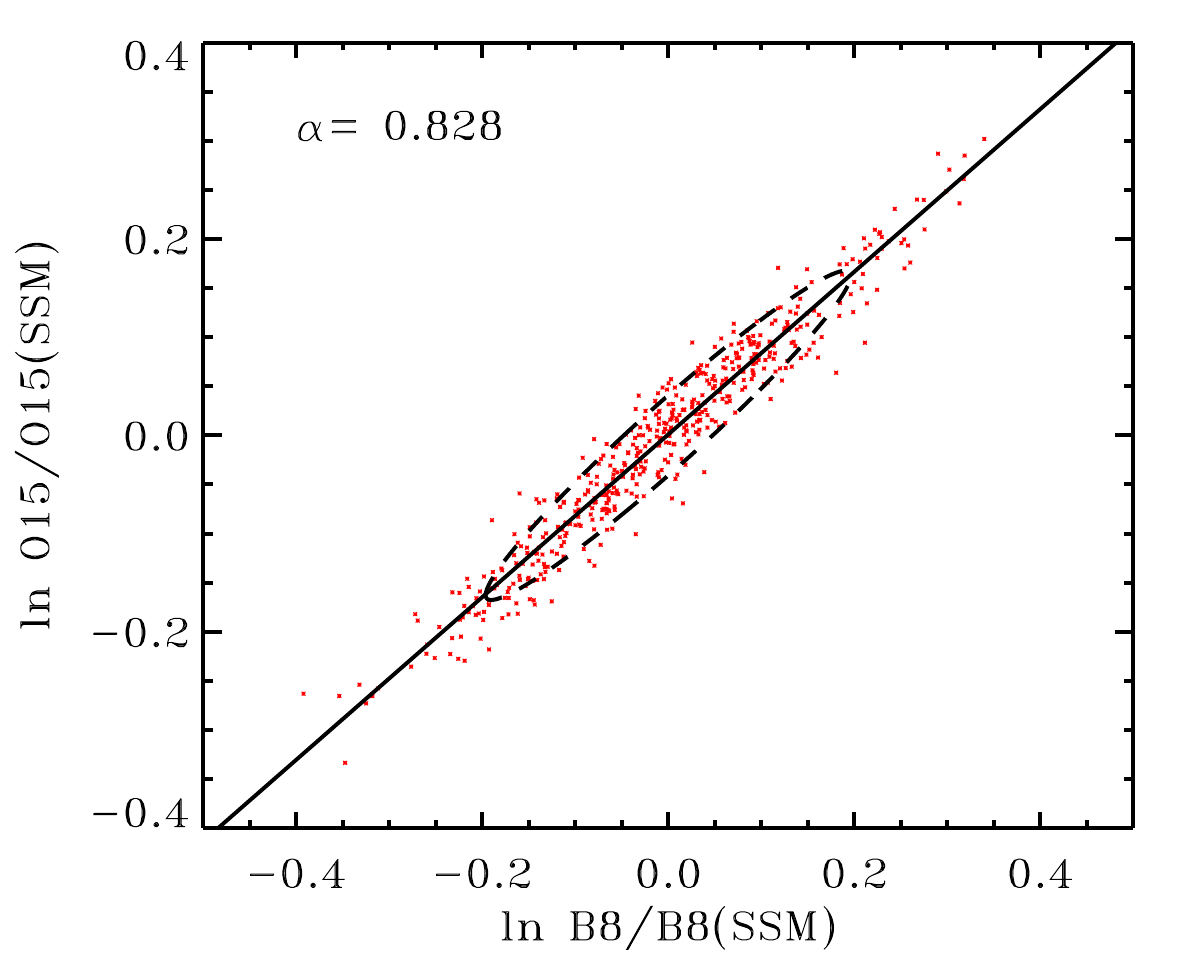}
\caption{The SSM $^8$B - $^{15}$O neutrino flux correlation, used in \cite{HS} to reduce SSM uncertainties.}
\end{figure}

Thus the current overall theoretical uncertainty in relating a future CN neutrino flux measurement
to core metalicity is about 9.6\%.  The dominate uncertainties, those due to flavor
physics and nuclear cross sections, can be reduced by future laboratory
measurements.  SNO+, a deep scintillator experiment that will be constructed in
SNOLab, may be able to measure the CN flux to an accuracy of about 10\% \cite{SNO+}.  Given
that recent changes in core metalicity are $\sim$ 30\%, it appears that future neutrino
experiments may be able to constrain core metalicity at an interesting level of precision.

   This work was supported in part by the Office of Nuclear Physics, U.S. Department of 
   Energy.

%
%
%

%
\end{document}